\documentclass[a4paper,11pt]{article}
\usepackage{pos}
\usepackage{ytableau}
\usepackage{xcolor}
\usepackage{amsmath}
\usepackage{multirow}

\title{Color structure of deuteron on the light front}

\author*[a,b]{Satvir Kaur}
\author[a,b]{Jiatong Wu}
\author[a,b,c]{Siqi Xu}
\author[a,b]{Chandan Mondal}
\author[a,b]{Xingbo Zhao}
\author[]{James P Vary$^c$ ~~~~~{\rm (BLFQ~Collaboration)}}

\affiliation[a]{Institute of Modern Physics, Chinese Academy of Sciences, Lanzhou, Gansu, 730000, China}

\affiliation[b]{School of Nuclear Physics, University of Chinese Academy of Sciences, Beijing, 100049, China}

\affiliation[c]{Department of Physics and Astronomy, Iowa State University, Ames, IA 50011, USA}

\emailAdd{satvir@impcas.ac.cn}
\emailAdd{wujt@impcas.ac.cn}
\emailAdd{xsq234@impcas.ac.cn}
\emailAdd{mondal@impcas.ac.cn}
\emailAdd{xbzhao@impcas.ac.cn}
\emailAdd{jvary@iastate.edu}

\abstract{We investigate the color structure of the deuteron by solving the light-front QCD Hamiltonian for its six-quark and six-quark–one-gluon components using basis light-front quantization. In this framework, the deuteron wavefunction consists of a singlet-singlet color state as well as additional hidden color states arising from non-trivial color rearrangements. Our analysis shows that while the singlet-singlet state is present, the hidden color states collectively dominate, contributing a larger probability to the deuteron wavefunction. These findings provide new insights into the role of hidden color components in the QCD description of nuclear structure.}

\FullConference{The XVIth Quark Confinement and the Hadron Spectrum Conference (QCHSC24)\\
 19-24 August, 2024\\
 Cairns Convention Centre, Cairns, Queensland, Australia\\}

\begin{document}

\maketitle

\section{Introduction}
The deuteron, as the simplest nuclear bound state, provides a unique testing ground for understanding the interplay between quantum chromodynamics (QCD) and nuclear structure. Traditionally, the deuteron has been described as a loosely bound system of a proton and a neutron, primarily governed by meson-exchange dynamics. However, from a fundamental QCD perspective, the deuteron is better understood predominantly a six-quark system, where color degrees of freedom play a fundamental role in its internal structure.

In QCD, quarks interact through the strong force, mediated by gluons, and governed by the SU(3) color symmetry. Each quark carries a color charge—red, green, or blue—while gluons facilitate interactions by exchanging color among them. Since isolated colored states cannot exist in nature because of color confinement, the quarks and gluons within a physically observed state interact to form an overall color-neutral state. Unlike baryons, which contain three quarks in a color-singlet configuration, the deuteron, as a six-quark system, admits more complex color structures, known as hidden-color states. The importance of hidden-color contributions to the deuteron has been highlighted in the literature, see for details~\cite{Matveev:1977xt, Matveev:1977ha, Hogaasen:1979qa, Brodsky:1983vf, Farrar:1991qi, Bashkanov:2013cla,Miller:2013hla, Bakker:2014cua}. While traditionally neglected in nuclear models, these states can contribute significantly to the deuteron wavefunction, particularly in observables related to short-range nuclear correlations and deep inelastic scattering.

A nucleon-based description of the deuteron is limited in its ability to capture these non-trivial color configurations. At high momentum transfers or short-distance scales, where quark and gluon degrees of freedom become relevant, effective nucleon-nucleon models break down. Investigating the deuteron at the quark level is therefore crucial for a more complete understanding of its internal structure and its role in high-energy nuclear reactions.

The basis light-front quantization (BLFQ) approach~\cite{Vary:2009gt} has been successfully implemented in various QCD systems, ranging from mesons to baryons, providing a non-perturbative framework to solve QCD from first principles. This approach has yielded significant insights into several hadronic properties by expanding the Fock space boundaries~\cite{Lan:2019vui, Lan:2019rba, Mondal:2019jdg, Lan:2019img, Lan:2021wok, Xu:2021wwj, Liu:2022fvl, Hu:2022ctr, Peng:2022lte, Xu:2022yxb, Zhu:2023lst, Zhu:2023nhl, Kaur:2023lun, Lin:2023ezw, Zhang:2023xfe, Kaur:2024iwn, Liu:2024umn, Yu:2024mxo, Nair:2024fit, Zhu:2024awq, Wu:2024hre, Lin:2024ijo, Xu:2024sjt, Peng:2024qpw, Lan:2025fia, Zhang:2025nll}. Given its success, we extend BLFQ to investigate the lightest nuclear bound state—the deuteron. This extension allows us to analyze the deuteron wavefunction directly in terms of quark and gluon degrees of freedom, providing a novel perspective on its color structure beyond the conventional nucleon-nucleon picture.

In this work, we employ the BLFQ approach to analyze the probability distribution of different color configurations within the deuteron wavefunction. To achieve this for an initial investigation, we truncate the Fock space to include only the six-quark ($qqq~qqq$) and six-quark–one-gluon ($qqq~qqq~g$) components. Specifically, we compute the probability of different color states in the deuteron wavefunction by solving the light-front QCD Hamiltonian in this truncated basis. This allows us to determine the relative contributions of the conventional nucleon-nucleon (singlet-singlet) configuration and the hidden-color states that emerge from non-trivial SU(3) color rearrangements. Our analysis reveals that hidden-color states, play a significant role in the full wavefunction.

\section{Formalism}
In BLFQ, we solve the eigenvalue problem of the light-front (LF) Hamiltonian: 
\begin{equation}
P^+ P^- |\Psi\rangle = M^2 |\Psi\rangle,
\end{equation}
where \( P^+ \) and \( P^- \) represent the longitudinal momentum and the LF Hamiltonian, respectively, acting on the deuteron state \( |\Psi\rangle \). This state is expanded in the Fock space to include various components of quarks (\( q \)), antiquarks (\( \bar{q} \)), and gluons (\( g \)) \cite{Brodsky:1997de}, such that  
\begin{equation}
|\Psi\rangle = \psi_{(qqq~qqq)} |qqq~qqq\rangle + \psi_{(qqq~qqq~g)} |qqq~qqq~g\rangle + \psi_{(qqq~qqq~q\bar{q})} |qqq~qqq~q\bar{q}\rangle + \dots
\end{equation}

where \( \psi(...) \) denotes the probability amplitudes of various partonic configurations \( |...\rangle \). In this work, we retain only the first two Fock components in Eq. (2), i.e., \( |qqq~qqq\rangle \) and \( |qqq~qqq~g\rangle \).
  
The LFQCD Hamiltonian with one dynamical gluon in the LF gauge (\( A^+ = 0 \)) is given by \cite{Brodsky:1997de}:  
\begin{align}
P^-_{\text{QCD}} = \int d^2x_{\perp} dx^- &\Big[  
\frac{1}{2} \bar{\psi} \gamma^+ \frac{m_0^2 + (i\partial^{\perp})^2}{i\partial^+} \psi  
+ A^\mu_a \frac{m_g^2 + (i\partial^{\perp})^2}{2} A_\mu^a  
+ g_s \bar{\psi} \gamma^\mu T^a A_\mu^a \psi  \nonumber\\
&
+ \frac{g_s^2}{2} \bar{\psi} \gamma^+ T^a \psi \frac{1}{(i\partial^+)^2} \bar{\psi} \gamma^+ T^a \psi  
\Big].
\end{align}

The first two terms in the above equation represent the contributions of kinetic energy, where \( m_0 \) and \( m_g \) are the bare masses of the quark and gluon, respectively. The fields \( \psi \) and \( A^\mu \) correspond to the quark and gluon fields. The coordinates \( x^- \) and \( x_{\perp} \) denote the longitudinal and transverse position coordinates, respectively. The remaining terms account for the interaction between partons within the meson, with \( g_s \) being the coupling constant. The matrices \( T^a \) are the generators of the SU(3) gauge group, while \( \gamma^\mu \) are the Dirac matrices.  

To incorporate quark mass corrections due to quantum fluctuations to higher Fock sectors, we introduce a mass counter term, \( \delta m_q = m_0 - m_q \), for the quark in the leading Fock component, where \( m_q \) is the renormalized quark mass \cite{Glazek:1993rc, Karmanov:2008br,Karmanov:2012aj, Zhao:2014hpa, Zhao:2020kuf}. Additionally, a mass parameter \( m_f \) is introduced to parameterize nonperturbative effects in the vertex interactions \cite{Glazek:1992aq, Burkardt:1998dd}. In our current Fock sector truncation, there are no mass corrections for the gluon, thus the mass counter term for the gluon is not needed, that is $m_g=0$.

In this framework, the longitudinal and transverse dynamics of Fock particles are described using a discretized plane-wave basis and 2D harmonic oscillator (HO) wave functions, respectively. The longitudinal motion is confined to a 1D box of length \(2L\), with antiperiodic (periodic) boundary conditions for fermions (bosons). The longitudinal momentum is given by \(p^+ = \frac{2\pi k}{L}\), where \(k\) takes half-integer (integer) values for fermions (bosons), and the zero mode for the boson is neglected. The total longitudinal momentum is \(P^+ = \frac{2\pi K}{L}\), where \(K = \sum_i k_i\), and the longitudinal momentum fraction of the \(i\)-th parton is \(x_i = \frac{p_i^+}{P^+} = \frac{k_i}{K}\).

For the transverse direction, the 2D HO wave function \(\phi_{n,m}(k_\perp; b)\) is used, with \(n\) and \(m\) representing the radial and angular momentum quantum numbers, and \(b\) being the HO scale parameter. The single-particle basis state is expressed by the set of quantum numbers \(\alpha = \{x, n, m, \lambda\}\), where \(\lambda\) denotes the helicity. The many-body basis state is the direct product of the basis states of single particles, \( |\alpha \rangle = \otimes_i |\alpha_i \rangle\), and the total angular momentum projection is defined as \(m_J = \sum_i (m_i + \lambda_i)\). The deuteron Fock sectors contain multiple color-singlet states, requiring additional labels to distinguish between different configurations. Notably, the first Fock sector contains five color-singlet states, while the second Fock sector contains sixteen color-singlet states.

The longitudinal and transverse truncations are controlled by two parameters: \(K\) and \(N_{\text{max}}\), respectively. The transverse truncation is set by \(N_{\text{max}} \geq 2n_i + |m_i| + 1\), while \(K\) determines the resolution in the longitudinal direction. \(N_{\text{max}}\) introduces ultraviolet (UV) and infrared (IR) cutoffs, \(\Lambda_{\text{UV}} = b\sqrt{N_{\text{max}}}\) and \(\Lambda_{\text{IR}} = \frac{b}{\sqrt{N_{\text{max}}}}\), respectively, regulating the high- and low-momentum regions.

All numerical calculations are performed with $N_{\text{max}} = 8$ and $K = 9$. The harmonic oscillator scale parameter is set to $b = 0.30$ GeV, while the UV cut off for the instantaneous interaction is $b_{\text{inst}} = 5.00$ GeV. The model parameters $\{m_u, m_d, m_f, g_s\} = \{1.00, 0.95, 42.56, 1.90\}$ (all in GeV except $g_s$) are determined by fitting the deuteron mass and electromagnetic properties. The large constituent quark masses in the first two Fock sectors partially account for confinement effects and contribute to the deuteron mass in a QCD bound state.

\section{Color states}
In this work, we describe the deuteron in Fock space with six quarks in the first Fock sector and six quarks plus one dynamical gluon in the second Fock sector. In SU(3) color symmetry, the decomposition of the six-quark Fock state follows the representation~\cite{Bakker:2014cua}:
\begin{align}
{\tiny \ydiagram{1} \otimes \ydiagram{1} \otimes \ydiagram{1} \otimes \ydiagram{1} \otimes \ydiagram{1} \otimes \ydiagram{1}}&= 
{\tiny \ydiagram{6} \oplus 5 ~~\ydiagram{5,1} \oplus 9 ~~\ydiagram{4,2} \oplus 10 ~~\ydiagram{4,1,1}} \nonumber\\
& {\tiny \oplus ~ 5 ~~\ydiagram{3,3} \oplus 16~~\ydiagram{3,2,1} \oplus 5~~\ydiagram{2,2,2}}
\end{align} 

\begin{align}
    \{ 3 \} \otimes \{ 3 \} \otimes \{ 3 \} \otimes \{ 3 \} \otimes \{ 3 \} \otimes \{ 3 \} &= 729 \nonumber\\ 
    & =\{28\}\oplus 5~ \{35 \} \oplus 9~ \{ 27 \} \oplus 10 ~\{10\} \oplus 5~ \{ 10 \} \oplus 16 ~\{8 \} \nonumber\\
    &\oplus 5~ \{1 \}
\end{align}

Among the five color-singlet states in the six-quark configuration, one is pure singlet from the singlet-singlet configuration, while the remaining four are hidden color states obtained from the octet-octet representation.

The possible color configurations in the second Fock sector of the deuteron, including six quarks and one gluon, are written as:
\begin{align}
{\tiny \ydiagram{1} \otimes \ydiagram{1} \otimes \ydiagram{1} \otimes \ydiagram{1} \otimes \ydiagram{1} \otimes \ydiagram{1} \otimes \ydiagram{2,1}} 
&= {\tiny \ydiagram{8,1} \oplus 6~~\ydiagram{7,2} \oplus 6~~\ydiagram{7,1,1}} \nonumber\\
&{\oplus \tiny ~ 14~~\ydiagram{6,3} \oplus  30~~\ydiagram{6,2,1} \oplus 14~~\ydiagram{5,4}} \nonumber\\
& {\tiny \oplus ~54~~\ydiagram{5,3,1} \oplus 40~~\ydiagram{5,2,2} \oplus 30~~\ydiagram{4,4,1} \oplus 61~~\ydiagram{4,3,2}}  \nonumber\\
&{\tiny \oplus~ 16 ~~\ydiagram{3,3,3}}
\end{align}

 \begin{align}
     \{ 3 \} \otimes \{ 3 \} \otimes \{ 3 \} \otimes \{ 3 \} \otimes \{ 3 \} \otimes \{ 3 \} \otimes \{8 \} =&~ 5832 \nonumber\\
     =& ~ \{80 \} \oplus 6 ~\{81 \} \oplus 6~ \{28\} \oplus 14~\{64\} \oplus 30~\{35\} \nonumber\\
     \oplus&~ 14~\{35\} \oplus 54~\{27\} \oplus 40~\{10 \} \oplus 30~\{10\} \nonumber\\
     \oplus&~ 61~\{8\} \oplus 16~\{1\}
 \end{align}
There exist 16 hidden color states in the second Fock sector of the deuteron. Therefore, in our case, the deuteron has 21 color states in total, of which one is a pure state, arising from the singlet-singlet combination of the two nucleons, while the rest are referred to as hidden color states. The detailed color configurations of the $qqq$-$qqq$ and $qqq$-$qqq$-$g$ systems responsible for the color-singlet states are presented in Table~\ref{Table:color}. 
The color factors, obtained by sandwiching the matrices $T^a$ and $T^aT^a$ between the color singlet wave functions, are used as multiplicative factors in the matrix elements of LF QCD Hamiltonian in the BLFQ basis to ensure the essential QCD feature in the deuteron calculation.


\begin{table}[hbt!]
\caption{The color singlet states corresponding to the different configurations in SU(3) color symmetry representation.}
\centering
\begin{tabular}{|l|l|c|c|c|}
  \hline
  & {\bf Color Configurations} & {\bf Total Color Singlet States} & \multicolumn{2}{c|}{\bf Probability (\%)} \\
  & & & \multicolumn{2}{c|}{(Preliminary)} \\
  \cline{4-5}
  & & & {\bf \(m_J = 0\)} & {\bf \(m_J = 1\)} \\ \hline
  \multirow{2}{*}{$| qqq~qqq \rangle$}  
  & Singlet-Singlet & 1 & 44.56 & 44.52 \\  
  & Octet-Octet & 4 & 12.98 & 13.02 \\ \hline
  \multirow{2}{*}{ } & Decuplet-Octet-Octet & 2 &  &  \\  
  & Octet-Decuplet-Octet & 2 &  &  \\  
  \multirow{3}{*}{$| qqq~qqq~g \rangle$} & Octet-Octet-Octet & 8 & 42.46 & 42.46 \\  
  & Octet-Singlet-Octet & 2 &  &  \\  
  & Singlet-Octet-Octet & 2 &  &  \\ \hline
\end{tabular}
\label{Table:color}
\end{table}

\begin{figure}[hbt!]
    \centering
    \includegraphics[width=0.8\linewidth]{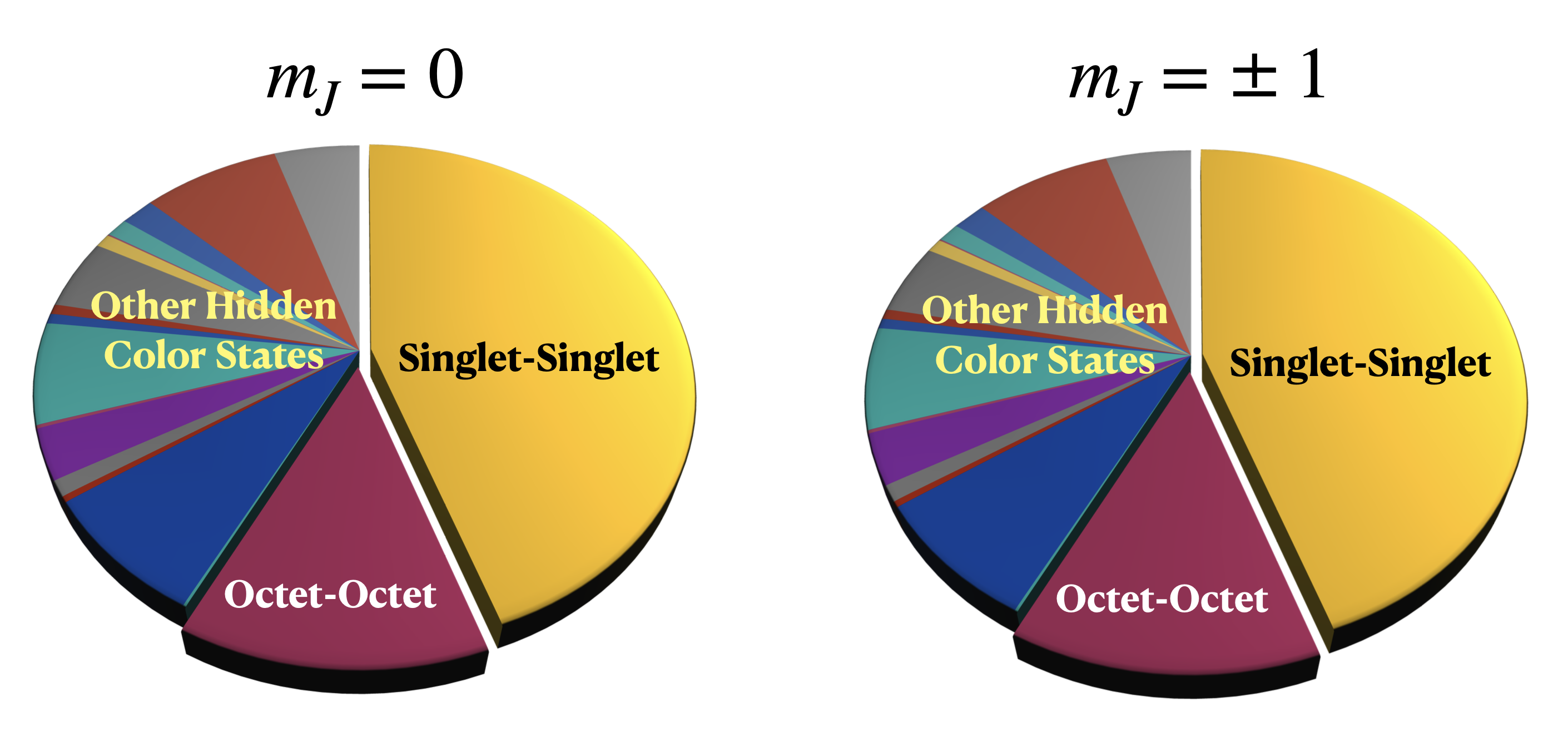}
    \caption{Probability (preliminary) of different color configurations within deuteron for $m_J=0$ and $m_J=\pm 1$ spin projections in our approach.}
    \label{fig:probability}
\end{figure}

Utilizing the light-front wave functions generated by diagonalizing the Hamiltonian, we compute the probability of the every color state to quantify the contribution of each color state.
As expected, our results show that the pure color state has a substantial probability at our model scale, while the hidden color states contribute the majority. Specifically, for both the components \( m_J = 0 \) and \( m_J =\pm 1 \), we find that the pure color state has a probability of approximately $44.5\%$, while the remaining $55.5\%$ of the probability is attributed to hidden color states, with only minor differences that are negligible, as shown in Table~\ref{Table:color}. To illustrate these contributions, we present probability distributions for different spin projections, specifically for \( m_J = 0 \) and \( m_J = \pm 1 \) in Fig.~\ref{fig:probability}.

\section{Conclusion}
In this work, we investigated the color structure of the deuteron using the basis light-front quantization framework. By solving the light-front QCD Hamiltonian within a truncated Fock space, we analyzed the probability distribution of different color configurations. Our results show that while the conventional singlet-singlet state contributes significantly, hidden color states dominate the deuteron wavefunction. This finding reinforces the idea that a purely nucleon-based description is insufficient to fully capture the internal structure of the deuteron.

\section*{Acknowledgement}
SK is supported by the Research Fund for International Young Scientists, 
Grant No.~12250410251, from the National Natural Science Foundation of China (NSFC), 
and the China Postdoctoral Science Foundation (CPSF), Grant No.~E339951SR0. 
CM is supported by new faculty start-up funding by the Institute of Modern Physics, 
Chinese Academy of Sciences, Grant No.~E129952YR0. 
XZ is supported by new faculty start-up funding by the Institute of Modern Physics, 
Chinese Academy of Sciences, by the Key Research Program of Frontier Sciences, 
Chinese Academy of Sciences, Grant No.~ZDBS-LY-7020, by the Natural Science Foundation 
of Gansu Province, China, Grant No.~20JR10RA067, by the Foundation for Key Talents 
of Gansu Province, by the Central Funds Guiding the Local Science and Technology 
Development of Gansu Province, Grant No.~22ZY1QA006, by the International Partnership 
Program of the Chinese Academy of Sciences, Grant No.~016GJHZ2022103FN, by the Strategic 
Priority Research Program of the Chinese Academy of Sciences, Grant No.~XDB34000000, 
and by the National Natural Science Foundation of China under Grant No.~12375143. 
JPV acknowledges partial support from the Department of Energy under Grant No.~DE-SC0023692. 
This research used resources of the National Energy Research Scientific Computing Center (NERSC), 
a U.S. Department of Energy Office of Science User Facility located at Lawrence Berkeley National Laboratory, 
operated under Contract No.~DE-AC02-05CH11231 using NERSC award NP-ERCAP0028672. 
A portion of the computational resources were also provided by the Gansu Computing Center. 
This research is supported by the Gansu International Collaboration and Talents Recruitment Base 
of Particle Physics, by the Senior Scientist Program funded by Gansu Province Grant No.~25RCKA008, and the International Partnership Program of the Chinese Academy of Sciences, 
Grant No.~016GJHZ2022103FN.

\end{document}